Comment

Evolution: Life has Evolved to Evolve

Comment on "How Life Changes Itself: The Read-Write (RW) Genome" by James Shapiro

Michael W. Deem
Departments of Bioengineering and Physics & Astronomy and
Graduate Program in Systems, Synthetic, and Physical Biology
Rice University
Houston, TX  77005 USA
mwdeem@rice.edu

Jim Shapiro synthesizes a great many observations about the mechanisms of evolution to reach the remarkable conclusion that large-scale modification, exchange, and rearrangement of the genome are common and should be viewed as fundamental features of life [1].  In other words, the genome should be viewed not as mostly read-only with a few rare mutations, but rather as a fully-fledged read-write library of genetic functions under continuous revision.  Revision of the genome occurs during cellular replication, during multicellular development, and during evolution of a population of individuals.  DNA formatting controls the timing and location of genetic rearrangements, gene expression, and genetic repair.  Each of these events is under the control of precise cellular circuits.  Shapiro reviews the toolbox of natural genetic engineering that provides the functionalities necessary for efficient long-term genome restructuring.

Shapiro quickly notes his preference for the term "DNA element" rather than "gene," due to the diversity of function that a given region of the genome may have, whether it be expression of protein, regulation of other regions of the genome, or scaffolding affecting the processing of the genome.  The classical theory of one gene, one protein is gone, replaced by an understanding that a given genetic region can be processed in multiple ways, leading to several possible functional results.  Indeed, the architecture of the genome is Lego-like, as pieces from distinct genomic regions can function together.  In prescient early work, Britten and Davidson speculated that distributed regulatory sites in the genome could form networks of control, which could also guide evolution of novel protein structure and function [2].  It is now known that there are *cis*-regulatory modules (CRMs) that encode multiple protein and RNA products that work together.  These modules coordinate the execution of routines that span the gamut from defining body parts during development to providing flexible responses for unpredictable contingencies that cells may face.  In eukaryotes the DNA is packaged around histones to form nucleosomes.  The formation of nucleosomes and their folding into higher-order chromatin complexes are regulated in ways that can affect replication, transcription, and expression.  There was even a hierarchy of symmetry breaking in the evolution of vertebrate gene regulatory elements: early evolution occurred in transcription factors and development genes, later evolution led to emergence of novel extracellular signaling genes, and recent evolution led to innovations in posttranslational protein modifiers [3].

Horizontal gene transfer (HGT) is one of the ways by which new DNA elements may be introduced to a genome.  In HGT, genetic material from a potentially different species can be introduced into the genome of an individual.  Viruses are particularly adept at copying genetic material, evolving it, and then inserting newly synthesized constructs into other species.  In this way, viruses serve as laboratories for evolutionary experimentation, performing a sort of synthetic biology on host genomes.  HGT is fundamental to evolution and occurs within and between all domains of life.  HGT is life's analog of the hierarchical strategies that have proven useful in protein molecular evolution [4].   There are now many specific examples of novel function created by HGT.  One of the mechanisms by which bacteria evolve antibiotic resistance is HGT of the enzymes that degrade antibiotics.  The adaptive immune system evolved by HGT of fragments that were then duplicated to form the VDJ elements and processing machinery [5].  Compelling evidence for domain accretion and domain shuffling in eukaryotic evolution was presented in the first draft sequencing of the human genome [6].  Many proteins have these mobile elements, providing support for Gilbert's exon-early theory of generation of the primordial fold diversity [7].  Viruses played a key role in the development of the placenta, facilitating the divergence of eutherian mammals from marsupials.  Indeed, viruses are an evolutionary melting pot: there are megaviruses containing sequences from all three domains of life.

Evolution within vertebrates is largely not due to a priori emergence of new proteins, the diversity of protein folds having been frozen in by the time vertebrates emerged, but rather to the evolution of new control strategies, such as *cis*-regulatory modules.  The mechanisms of natural genetic engineering conspired to create novel regulatory functions, with at least 10-20% of the new regulatory elements coming from mobile genetic elements.  These mobile genetic elements are so abundant that there are mechanisms to control and regulate them: CRISPR in bacteria, piRNA in animals, and siRNA in plants.

Duplication of genetic elements and splicing of disparate pieces of genetic material are fairly common processes in the genome.  Joining of broken ends is a complex process involving marking by histones, transportation to a special repair region of the nucleus, and finally joining and ligation by a multi-protein non-homologous end-joining complex.  Much chromosome diversity has arisen from duplication and rejoining of genetic material.  Occasionally, the duplicated material is even inserted in an inversed orientation.  Shapiro argues cogently that this sophisticated machinery has likely evolved just for this process of diversity generation, rather than as the byproduct of some other process [1].  An example of functional diversity arising from duplication is the extensive diversity of olfactory receptors in mammals. Duplication of genetic elements also led to the diversity of splice variants that give different hair cells exquisite frequency range sensitivity [8].

To make his point, Shapiro mentions several fantastic examples of extensive DNA rearrangement on the timescale of a single generation [1].  For example, ciliates, a group of protozoans, possess a micronucleus and a macronucleus in each individual. The DNA in the micronucleus is fragmented and randomly recombined into the macronucleus during the life cycle of a ciliate.  Hundreds to thousands of rearrangements can occur during this extreme example of exon shuffling.  Another example occurs in various pathogens that can extensively rearrange the proteins visible to the host immune system.  This 'phase variation' allows a pathogen to continually evade destruction by the immune response and to survive as a population.  A final example is found in developing neurons, where a DNA retrotransposon can cause cut-and-paste duplications that lead to extensive neuronal diversity in a single developing individual [9].

Hybridization is an example of evolution by large-scale genetic manipulation.  Hybridization is mating between individuals from different species or populations.  Hybridization is a common mechanism to generate diversity and new species of plants.  Shapiro points out that the extensive observed seed plant diversity, which bothered Darwin as not easily explainable from slow accumulation of single amino acid changes, is the result of hybridization.

Shapiro, thus, teaches that the genome is profitably viewed as a collection of genetic subroutines that may be mixed, matched, duplicated, and joined together to create novel functional diversity [1]. He argues that combination of functional components is likely a better strategy than is a random, unbiased search of sequence, an assertion borne out by simulations of protein molecular evolution [4]. He points out that the space of possibilities is smaller, and thus evolution is faster, if functional parts are used [1], a result previously noted in models for the emergence of modularity in biology [10,11]. This large-scale restructuring of the genome occurs in response to stress that may occur during the natural cell life cycle, due to DNA damage, or in response to severe ecological challenge. Shapiro concludes that these natural genetic engineering mechanisms have evolved to facilitate adaption of life to the turbulent history of the planet [1], i.e. that life has evolved to evolve [12].